\documentclass[twocolumn,aps,pra,amsmath,amssymb,crop,tikz]{revtex4-2}
\usepackage[T1]{fontenc}
\usepackage{graphicx}
\usepackage{babel}
\usepackage{braket}
\usepackage{qcircuit}
\usepackage{xcolor}
\usepackage{dcolumn}
\usepackage{centernot}
\usepackage{mathtools}
\usepackage{stmaryrd}

\usepackage[colorlinks,citecolor=red,urlcolor=blue,bookmarks=false,hypertexnames=true]{hyperref} 

\usepackage{fancyhdr}

\begin{document}

\title{Implementing Jastrow--Gutzwiller operators on a quantum computer using the cascaded variational quantum eigensolver algorithm}

\author{John P. T. Stenger}
\affiliation{U.S. Naval Research Laboratory, Washington, DC 20375, United States}
\author{C. Stephen Hellberg}
\affiliation{U.S. Naval Research Laboratory, Washington, DC 20375, United States}
\author{Daniel Gunlycke}
\affiliation{U.S. Naval Research Laboratory, Washington, DC 20375, United States}

\begin{abstract}
A Jastrow--Gutzwiller operator adds many-body correlations to a quantum state.  However, the operator is non-unitary, making it difficult to implement directly on a quantum computer.  We present a novel implementation of the Jastrow--Gutzwiller operator using the cascaded variational quantum eigensolver algorithm.  We demonstrate the method on IBM Q Lagos for a Hubbard model.
\end{abstract}

\maketitle
\thispagestyle{fancy}

\section{Introduction}

Quantum computers access exponentially large Hilbert spaces that cannot be efficiently simulated using classical computing~\cite{Feynman1982,Benioff1980}.  Significant effort has been invested into using quantum computers to determine the ground state of quantum systems~\cite{Lanyon2010,Wecker2015}.  These algorithms 
can be categorized into two types: those that 
time evolve the quantum system (e.g., the quantum phase estimation algorithm~\cite{kitaev1995} and probe qubit algorithms~\cite{stenger2022,Wang2012,Terhal2000}) and those that evaluate expectation values of the Hamiltonian with a variational ansatz state (e.g., the variational quantum eigensolver algorithm~\cite{Peruzzo2014,Kandala2017,McArdle2020,McClean2016,Cerezo2021,OMalley2016,Wang2019,Arute2020,Gonthier2020,HeadMarsden2021}).  Variational algorithms usually require shorter quantum circuits than those required by time-evolution algorithms.  Having short quantum circuits is important on current and near-term quantum computers as these devices can only perform a limited number of gate operations before the quantum state decoheres.  Therefore, variational algorithms are likely to be among the first to show quantum advantage.

Variational algorithms require that a particular ansatz is chosen for the form of the ground state.  
Examples include the unitary coupled cluster ansatz~\cite{Peruzzo2014,Hempel2018,Romero2018,Harsha2018,Bauman2020,Dallaire2019,Sokolov2020,Yunseong2020,Lee2019,Grimsley2020,Motta2021,Matsuzawa2020,Kivlichan2018,Setia2019}, the quantum alternating operator ansatz~\cite{Farhi2014,Hadfield2019,Lloyd2018,Morales2020,Wang2020}, the variational hamiltonian ansatz~\cite{Wecker2015,Wiersema2020,Ho2019}, and the hardware-efficient ansatz~\cite{Kandala2017,Kandala2019,Ganzhorn2019,Barkoutsos2018,Gard2020,Otten2019,Tkachenko2021,Kokail2019,Bravo_Prieto2020}.  There is no ansatz which works best in every case because the space of quantum states is exponentially large while the ansatz generally has a subexponential number of parameters.  The ansatz must be chosen such that its manifold includes a path connecting the initial state to a good approximation of the ground state of the particular Hamiltonian being considered.  Therefore, it is important to have a variety of ansatzes with some intuition as to when each should be applied.  A powerful approach for improving variational states is to use a Jastrow--Gutzwiller operator to correlate the electrons in a trial many-body state~\cite{Jastrow1955,Gutzwiller1963,Brinkman1970,Vollhardt1984,Vollhardt1984b,Hellberg1991,Hellberg2000,Capello2005,Capello2006,Edegger2007,Fabrizio2007,Haldane1988,Shastry1988,Yunoki2006,Tahara2008,LeBlanc2015,Kandala2017,Ferrari2022}.  The Jastrow operator rescales the quantum state using density-density correlations~\cite{Jastrow1955}.  The Gutzwiller operator is a specific form of the Jastrow operator in which only doubly occupied orbitals are included~\cite{Gutzwiller1963}.  Because the Jastrow--Gutzwiller operator is non-unitary, implementing it on a quantum computer is challenging~\cite{Seki2022,Mazzola2019,Yao2021,Murta2021}.

In this paper, we apply the Jastrow--Gutzwiller operator using the cascaded variational quantum eigensolver (CVQE) algorithm~\cite{Gunlycke}.  The ansatz in the CVQE algorithm is 
a product of a diagonal operator dependent on the variational parameters and a unitary operator.  This includes the special case $\hat{A}(\theta) = \hat{G}(\theta)\hat U$, where $\theta$ is a collection of variational parameters, $\hat{G}(\theta)$ is the non-unitary Jastrow--Gutzwiller operator, and $\hat U$ is a unitary Thouless operator.  We show how to use this ansatz to perform calculations for an arbitrary fermionic system.  As an illustration, we demonstrate our method for the four-site square and triangular Hubbard models with periodic boundary conditions.

Section~\ref{s.3} presents specific forms of the operators $\hat{G}(\theta)$ and $\hat{U}$, the 
derivation of the expectation values for a general Hamiltonian $\hat H$, and the mapping to a quantum computer.  In Sec.~\ref{s.5}, we demonstrate the method for the Hubbard model.  In Section~\ref{s.6} presents results of the demonstration obtained in part from calculations performed on IBM Q Lagos for the Hubbard model on square and triangular lattices~\cite{Dagotto1994,Imada1998,Hellberg1999,Cao2018}.  Conclusions are shared in Section~\ref{s.7}.

\section{Method}
\label{s.3}
Consider a general fermionic system and let $\mathcal{Q}=\{0,1,...,Q-1\}$ be an index set for the basis states $\ket{\phi_q}$ for the $Q$-dimensional one-body Hilbert space.  A basis for the many-body Hilbert space can then be formed by Fock states $\ket{\Phi_n}$ identified by fermionic configurations $n=(n_q)_{q\in\mathcal{Q}}$ of occupation numbers in $\{0,1\}$, where $n_q$ equals $0$ and $1$ mean that the one-body state $\ket{\phi_q}$ is unoccupied and occupied, respectively.

We prepare the initial state to be one of these Fock states, which we can express as 
\begin{equation}
   \ket{\Phi_n} = \prod_{q\in\mathcal{Q}}\big(c^{\dagger}_q\big)^{n_q} \ket{0},
   \label{e.7}
\end{equation}
where $c_{q}^{\dagger}$ is the fermionic creation operator associated with $q$ and $\ket{0}$ is the vacuum state.

\subsection{The Thouless operator}
\label{TTO}

We are often interested 
in a state
\begin{equation}
    \ket{\Psi_n}= \hat{U}\ket{\Phi_n},
     \label{Uc}
\end{equation} 
which is not necessarily a Fock state based on the initial one-body basis $\{\ket{\phi_q}\}$.  
In our example in Section~\ref{s.6}, $ \ket{\Psi_n}$ will be the Fermi sea~\cite{Gros1989}.
We consider states that can be produced by transforming the one-body Hilbert space using the unitary change-of-basis operator $\hat f$ defined by
\begin{equation}
	\ket{\psi_q}=\sum_{q'\in\mathcal{Q}}f_{qq'}\ket{\phi_{q'}},
   \label{e.8}
\end{equation}
where $f_{qq'}=\braket{\phi_{q'}|\psi_q}$ are complex coefficients.  
In Appendix~\ref{GSD}, we show that the effect of this transformation on the many-body Fock space is captured by the Thouless operator~\cite{Thouless1960,Murta2021,Wecker2015,Kivlichan2018,Jiang2018,Somma2002}
\begin{equation}
    \hat{U} = \exp\Big[\sum_{q,q'\in\mathcal{Q}}\big(\log\hat f\big)_{qq'}c^{\dagger}_{q'} c_q\Big],
   \label{thouless}
\end{equation}
where $\log\hat f$ is the operator logarithm of $\hat f$, which is guaranteed to exist as $\hat f$ is invertible.

We implement the unitary operator $\hat U=\hat U(\hat f)$ in a quantum circuit by decomposing the one-body operator $\hat f$ into a set of elementary operators $\hat{f}_m$ such that $\hat f=\hat{f}_1\hat{f}_2\cdots\hat{f}_M$.  The method for generating these transformations and the associated quantum circuit is presented in Appendix~\ref{CMFTTO}\@.  It then follows that
\begin{equation}
    \hat{U}(\hat{f}) = \hat{U}(\hat{f}_1)\,\hat{U}(\hat{f}_2)\cdots\hat{U}(\hat{f}_m),
    \label{Uff}
\end{equation}
where each $\hat{U}(\hat{f}_m)$ is 
implemented using standard one- and two-qubit gates.

\subsection{The Jastrow--Gutzwiller operator}
\label{TTO}

To address the effects of interactions, we use the Jastrow--Gutzwiller operator~\cite{Jastrow1955,Gutzwiller1963}, which can be expressed as
\begin{equation}
    \hat{G}(\theta) = \exp \bigg( - \sum_{q q'} \theta_{q q'} \hat{n}_q \hat{n}_{q'} \bigg),
    \label{J-G}
\end{equation}
where $\hat{n}_{q} = c^{\dagger}_qc_q$ is the number operator for orbital $q$, and $\theta_{qq'}$ is a real-valued variational parameter in the collection $\theta=(\theta_{qq'})_{qq'\in\mathcal Q}$.

The energy expectation value in the CVQE algorithm takes the form~\cite{Gunlycke}
\begin{equation}
    E(\theta) = \frac{\bra{\Psi_n}\hat{G}(\theta)\hat H\hat{G}(\theta)\ket{\Psi_n}}{\bra{\Psi_n}\hat{G}^2(\theta)\ket{\Psi_n}},
\end{equation}
where $\hat H$ is the Hamiltonian of the system.  The denominator is necessary because $\hat{G}$ is a non-unitary operator.  As we will show below, both the numerator and the denominator can be computed efficiently.  

The full ansatz $\hat{A}(\theta) = \hat{G}(\theta) \hat{U}$ is particularly useful  when we have a single-particle solution defined by $\hat{f}$ (e.g., the eigenstates of the kinetic term of the Hamiltonian), and we want to turn on certain interactions defined by a diagonal operator~\cite{Gunlycke}.  
In Sec.~\ref{s.5} we demonstrate the ansatz with a Hubbard model.

Let the Hamiltonian be a general linear combination of products of fermion operators
\begin{equation}
    \hat H = \sum_{l=0}^{N_H-1} h_l \hat{N}_{\mathcal Q^N_l} C^{\dagger}_{\mathcal Q^+_l} C_{\mathcal Q^-_l}
\end{equation}
where $N_H$ is the number of terms in the Hamiltonian, $h_l$ are defining parameters, $\hat{N}_{\mathcal Q'}=\prod_{q \in \mathcal Q'}\hat{n}_q$, $C_{\mathcal Q'}^\dagger=\prod_{q \in \mathcal Q'}c_q^\dagger$, for any index set $\mathcal Q'$, and $\mathcal Q^N_l$, $\mathcal Q^+_l$ and $\mathcal Q^-_l$ are disjoint subsets of $\mathcal Q$.  Let us also define $\mathcal Q_l=\mathcal Q_l^+ \cup \mathcal Q_l^-$ and $\bar{\mathcal Q}_l = \mathcal Q \backslash \mathcal Q_l $.  Using this form of the Hamiltonian, we obtain
\begin{align}
    \hat{G}(\theta)\hat H\hat{G}(\theta) &= \sum_{l=0}^{N_H-1} h_l \hat{N}_{\mathcal Q^N_l} C^{\dagger}_{\mathcal Q^+_l} C_{\mathcal Q^-_l}\nonumber\\
    &\times \exp\!\bigg(\!-\sum_{q q'}\left[ \epsilon_{lq} + \epsilon_{lq'} + \zeta_{lqq'}\right] \theta_{qq'}\hat{n}_q^{\epsilon_{lq}}\hat{n}_{q'}^{\epsilon_{lq'}}\bigg)
\label{OHOf}
\end{align}
where the binary variables
\begin{equation}
    \epsilon_{lq} = \begin{cases}
        1 & \text{if } q \in \bar{\mathcal Q}_l,\\
        0 & \text{if } q \in \mathcal Q_l,
    \end{cases}
\label{elq}
\end{equation}
and 
\begin{equation}
    \zeta_{lqq'} = \begin{cases}
        1 & \text{if } q,q' \in \bar{\mathcal Q}^+_l,\\
        1 & \text{if } q,q' \in \bar{\mathcal Q}^-_l,\\
        0 & \text{ otherwise } .
    \end{cases}
\label{zlq}
\end{equation}
The derivation is given in Appendix~\ref{do}\@.  
We see in Eq.~(\ref{OHOf}) that the creation and annihilation operators never act on the same orbital as the number operators.  Because the number operators are diagonal in the occupation basis, we do not need separate measurements for each term in the exponential.  

In order to perform measurements of $\hat{G}(\theta)\hat H\hat{G}(\theta)$ on a quantum computer we need to map the fermion operators onto qubit operators.  Since we require that the number operators are diagonal, a natural map is the Jordan--Wigner transformation 
\begin{equation}
\begin{split}
    c^{\dagger}_{q} = \frac{X_{q} - i Y_{q}}{2} \prod_{q'=0}^{q-1}Z_{q'},
\end{split}
\label{J-W}
\end{equation}
where $X_q$, $Y_q$, and $Z_q$ are the Pauli-x, Pauli-y, and Pauli-z operators respectively.  Once the fermion operators are replaced with Pauli operators, we perform measurements of $\hat{G}(\theta)\hat H\hat{G}(\theta)$ on the quantum computer in the same way one would perform measurements of a Hamiltonian, except that we do not expand the exponential in terms of Pauli strings.  Instead, we take measurements of the qubits in $\bar{\mathcal Q}_l$  and insert those values directly into the exponential function.  This is possible because all of the Pauli terms in the exponential can be measured in the same basis.  See appendix~\ref{do} for more information on the measurement procedure.

\section{The Hubbard Model with the Gutzwiller ansatz}
\label{s.5}

We select the Hubbard model as a test case to demonstrate our method.  The Hubbard model is a particularly useful model for near-term quantum computing as it is simple enough to have relatively few terms yet complex enough that it has not been solved classically except for special cases.  Furthermore, it is a physically important model as it can be used to describe the most relevant interactions in a variety of physical systems.   The Hamiltonian for the Hubbard model is
\begin{equation}
    \hat H = \mu \hat{M} + k \hat{K} + d \hat{D},
\end{equation}
where $\mu$ is the chemical potential, 
\begin{equation}
    \hat{M} = \sum_{i\sigma}\hat{n}_{i\sigma},
\end{equation}
$k$ is a kinetic energy parameter,
\begin{equation}
     \hat{K} = \sum_{\langle i,j\rangle}\sum_\sigma \left( c^{\dagger}_{i\sigma}c_{j\sigma} + c^{\dagger}_{j\sigma}c_{i\sigma} \right), 
\end{equation}
$d$ is the interaction strength, and
\begin{equation}
    \hat{D} = \sum_i \hat{n}_{i\uparrow}\hat{n}_{i\downarrow},
\end{equation}
where $c^{\dagger}_{i\sigma}$ creates an electron on site $i\in\{0,1,...,N-1\}$ with spin $\sigma\in\{\uparrow,\downarrow\}$, $\hat{n}_{i\sigma} = c^{\dagger}_{i\sigma}c_{i\sigma}$ is the number operator, $N$ is the number of sites, and $\langle i,j\rangle$ denotes neighboring sites $i$ and $j$. 

Writing each term 
as a summation over Pauli strings using the Jordan--Wigner transformation in Eq.~\eqref{J-W}, the Hamiltonian operators become
\begin{equation}
    \hat{M} = \sum_{i\sigma} \frac{1-Z_{i\sigma}}{2},
\end{equation}
\begin{equation}
    \hat{D} = \sum_{i} \frac{1-Z_{i\uparrow}}{2} \frac{1-Z_{i\downarrow}}{2},
\end{equation}
and
\begin{equation}
     \hat{K} =\sum_{\langle i,j\rangle}\sum_{\sigma} \bigg( X_{i\sigma}\prod_{l=i+1}^{j-1} Z_{l\sigma}X_{j\sigma}+ Y_{i\sigma}\prod_{l=i+1}^{j-1} Z_{l\sigma}Y_{j\sigma} \bigg).
\label{KJW}
\end{equation}

In this demonstration, we apply the Gutzwiller operator 
\begin{equation}
     \hat{G}(\theta) = e^{-\theta \hat{D}}.
\end{equation}
for the non-unitary factor of the ansatz.  Notice that there is only a single optimization parameter in this case.  

We next determine the Pauli operators that need to be computed with the quantum computer.  
Both $\hat{M}$ and $\hat{D}$ contain only Pauli-$z$ operators and commute with $\hat{G}(\theta)$.  
Therefore, we have $\hat{G}(\theta)\hat{M}\hat{G}(\theta) = \hat{M}\hat{G}^2(\theta)$ and $\hat{G}(\theta)\hat{D}\hat{G}(\theta) = \hat{D}\hat{G}^2(\theta)$, which are readily computable on the quantum computer.  In contrast, $\hat{K}$ contains Pauli-$x$ and Pauli-$y$ operators that need to be factored so that each term in the expectation value can be easily diagonalized.  The operator can be written as
\begin{equation}
\begin{split}
    &\hat{G}(\theta)\hat{K}\hat{G}(\theta) =
     \\
     &\sum_{\langle i,j\rangle}\sum_{\sigma}  \left( X_{i\sigma} \prod_{l=i+1}^{j-1} Z_{l\sigma} X_{j\sigma} + Y_{i\sigma} \prod_{l=i+1}^{j-1} Z_{l\sigma} Y_{j\sigma} \right)
     \\
     & \times  e^{-\frac{\theta}{2}(1-Z_{i\bar{\sigma}})}e^{-\frac{\theta}{2}(1-Z_{j\bar{\sigma}})}e^{-\frac{\theta}{2} \sum_{k \neq i,j} (1-Z_{k\uparrow})(1-Z_{k\downarrow})}  .   
\end{split}
\label{GKG}
\end{equation}
where $\bar\sigma$ is the complement of $\sigma$ (i.e., the opposite spin).  For each term in the summation, the Pauli-$x$ and Pauli-$y$ operators have been factored such that there is at most one operator acting on any particular spin-orbital.  This allows us to easily diagonalize each term in the summation so that they can be measured separately on the quantum computer. See Appendix~\ref{DOGKG} for a derivation of Eq.\,(\ref{GKG}).  

All of the Pauli-$z$ operators can be measured simultaneously.  Both $\braket{\hat{G}(\theta)\hat{D}\hat{G}(\theta)}$ and $\braket{\hat{G}(\theta)\hat{M}\hat{G}(\theta)}$ can be measured with a single run of the quantum computer (where a run is composed of enough shots to achieve statistically accurate measurements). The expectation value  $\braket{\hat{G}(\theta)\hat{K}\hat{G}(\theta)}$ requires a run for each term in the summation, which is proportional to the number of lattice sites. 

\section{demonstration results}
\label{s.6}

\begin{figure}[h]
\vspace{2mm}
\begin{center}
\includegraphics[width=\columnwidth]{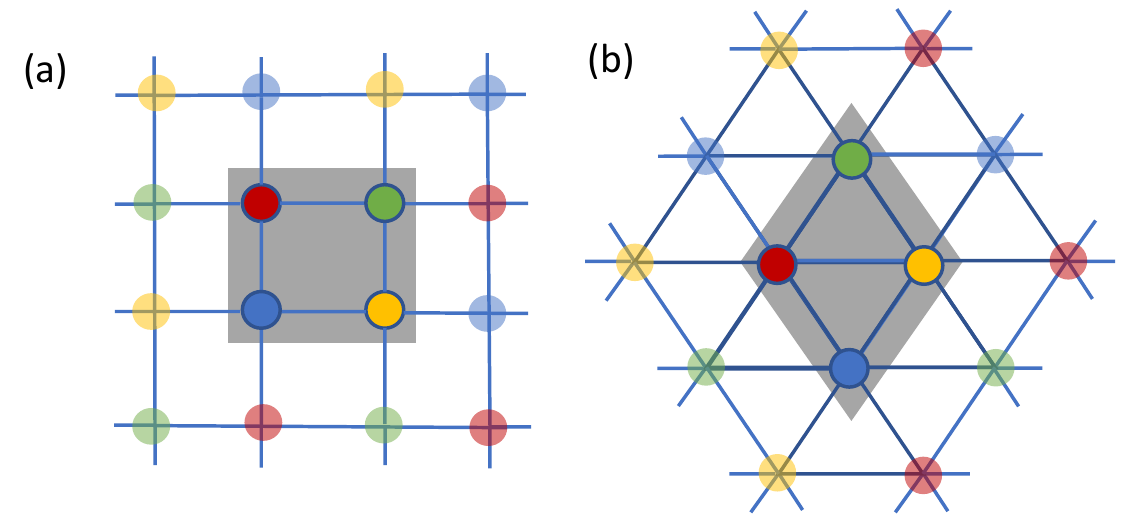}
\end{center}
\vspace{-2mm}
\caption{
Lattices used to demonstrate the algorithm:  (a) A four-site square lattice with periodic boundary conditions. (b) A four-site triangular lattice with periodic boundary conditions.  The solid circles indicate the four sites while the transparent circles are periodic images.  
}
\label{lattice}
\vspace{-3mm}
\end{figure}

\begin{figure}[h]
\vspace{2mm}
\begin{center}
\includegraphics[width=\columnwidth]{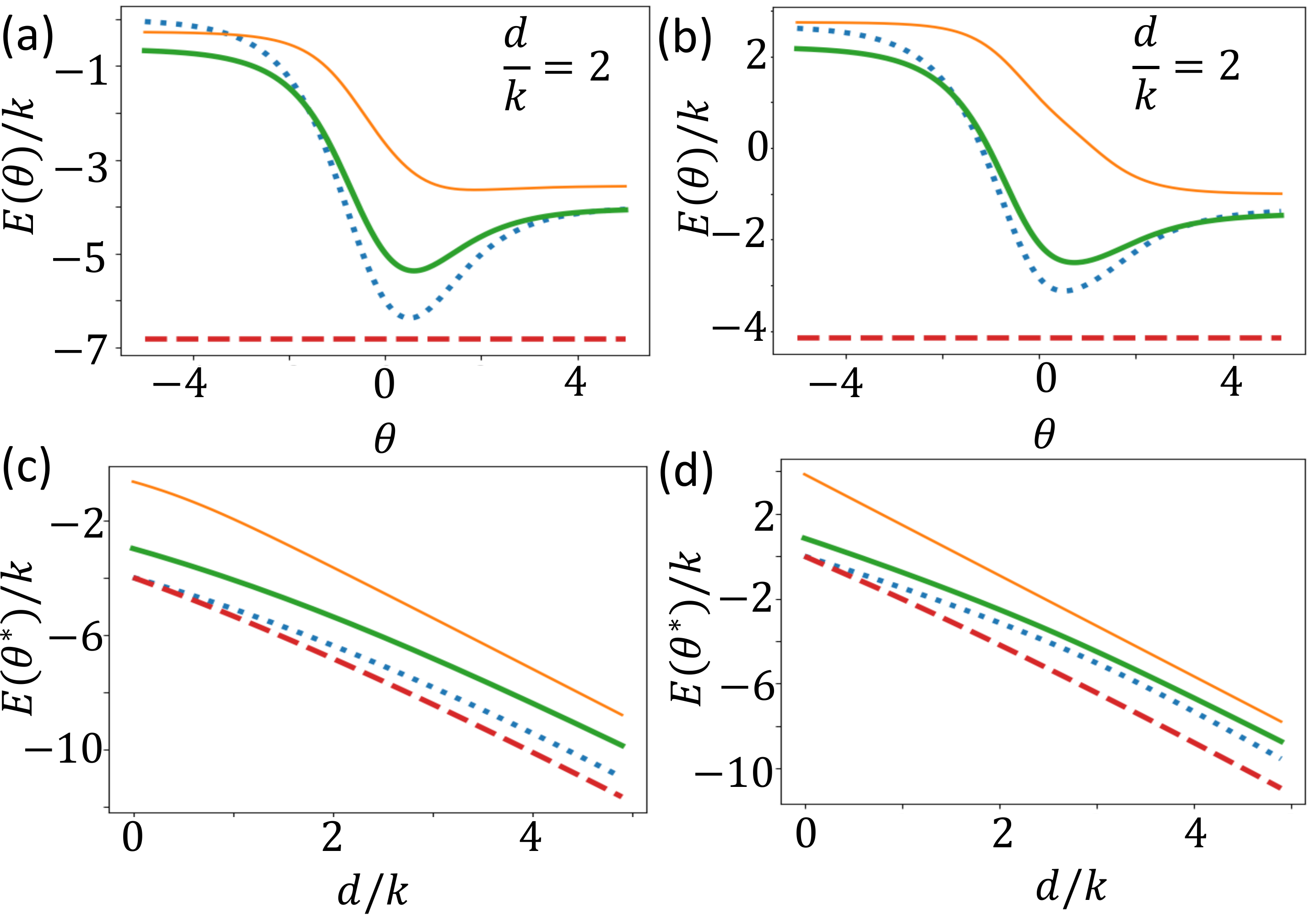}
\end{center}
\vspace{-2mm}
\caption{Ground-state energy for the square and triangular lattices.  The dashed-red line is the exact energy calculated by diagonalizing the Hamiltonian. The dotted-blue curve is an exact classical simulation of the Gutzwiller approach, the thick-green curve is the Gutzwiller approach using the data from the
IBM Q Lagos quantum computer for spin-down electrons only, and the thin-orange curve uses the quantum computer for both the spin-up and the spin-down expectation values.  (a) and (b) show the expectation value of the energy $E(\theta)$ as a function of the Gutzwiller parameter $\theta$.  (a) is for the square lattice and (b) is for the triangular lattice.  The optimal value $\theta^*$ of $\theta$ is found by locating the minimum of $E(\theta)$.  (c) and (d) show the optimal expectation value $E(\theta^*)$ as a function of the interaction strength 
$d$ for the square lattice and the triangular lattice, respectively.  For information on the specifications of IMB Q Lagos, see Appendix~\ref{DS}.}
\label{energies}
\vspace{-3mm}
\end{figure}

We demonstrate our algorithm on a four-site square lattice and a four-site triangular lattice both with periodic boundary conditions.  The calculations are performed at the center point $\Gamma$ in the first Brillouin zone, and thus, no phases are applied to the hopping terms across the periodic boundaries~\cite{Lin2001}.
The two lattices are represented in Fig.~\ref{lattice}.  The chemical potential is set so that the ground state has half filling (two spin-up electrons and two spin-down electrons).  For the square and triangular lattices, we set $\mu=-d/2$ and $\mu=1-2d/3$, respectively.  In order to access the half-filled state on the quantum computer we apply the two-particle Slater determinant circuit as described in Sec.~\ref{TTO}.  

We vary the single parameter $\theta$ to minimize the expectation value of the energy.
By moving the evaluation of $\hat{G}(\theta)$ to the classical computer, we have not only removed the need to rerun the quantum circuits when we vary the parameters, but we have also removed the coupling between the spin species from the quantum computer.  
Evaluating the circuit for only a single spin species reduces the number of required qubits by half~\cite{Stenger2022b}.

Fig.~\ref{energies}(a,b) shows the energy expectation values for the square and triangular lattices as a function of $\theta$ for $d=2k$.  The dashed red-line shows the exact ground-state energy which is plotted for reference.  The dotted-blue curve shows the energy expectation value for a noise-free simulation of our method.  

The two solid curves are computed using the quantum device.  
 The thin-orange curve is a direct calculation of all electrons on the quantum computer. 
 The thick-green curve is calculated using the quantum computer for the up electrons and the noise-free simulator for the down electrons.
The curves are smooth even though the data is noisy because $\theta$ is varied using the classical computer after the quantum measurement have been recorded. The minimum energy expectation value can occur at infinite $\theta$ as in the thin-orange curve in Fig.~\ref{energies}b.  

Figs.~\ref{energies}(c,d) shows the optimized energy expectation values 
as a function of $d$.  At $d=0$ the Gutzwiller state is the exact ground state.  Noise in the quantum circuit generates error in the expectation values.  The error is greatly reduced when we run only a single spin species on the quantum computer.  
Furthermore, the error in the energy expectation value stays relatively constant as a function of $d$ and even slightly decreases when $d\approx k$.

\section{conclusion}
\label{s.7}
We have proposed a method for implementing a generalized Jastrow--Gutzwiller ansatz on a quantum computer using the cascaded variational quantum eigensolver algorithm.
This allows us to implement the Jastrow--Gutzwiller ansatz without adding auxiliary qubits or making approximations to the operator.  We demonstrate the method on the IBM Q Lagos device for a four-site Hubbard model on both square and triangular lattices at half-filling. 

\begin{acknowledgments}

This work has been supported by the Office of Naval Research (ONR) through the U.S. Naval Research Laboratory (NRL).  J.P.T.S. thanks the National Research Council Research Associateship Programs for support during his postdoctoral tenure at NRL.  We acknowledge quantum computing resources from IBM through a collaboration with the Air Force Research Laboratory (AFRL).

\end{acknowledgments}

\appendix

\section{Generating Slater Determinants}
\label{GSD}

In this appendix, we derive Eq.~\eqref{Uc}, where $\hat{U}$ is given by Eq.~\eqref{thouless} and $c^{\dagger}_q$ and $a^{\dagger}_q$ are related by
\begin{equation}
    a^{\dagger}_{q} = \sum_{q'} \hat{f}_{qq'}c^{\dagger}_{q'}.
\end{equation}
Our derivation follows closely the one given in~\cite{Kivlichan2018}.  Specifically, we show that
\begin{equation}
    \hat{U} c^{\dagger}_q\hat{U}^{\dagger} = a^{\dagger}_q.
\end{equation}
To do so, let us define $\hat{\kappa} = \sum_{q q'}\log(\hat{f})_{q q'}c^{\dagger}_{q'} c_q $.  Using the Baker--Campbell--Hausdorff formula we have
\begin{equation}
    \hat{U} c^{\dagger}_q \hat{U}^{\dagger} = e^{\hat{\kappa}}c^{\dagger}_q e^{-\hat{\kappa}} = c^{\dagger}_{q} + [\hat{\kappa},c^{\dagger}_q] + \frac{1}{2}[\hat{\kappa},[\hat{\kappa},c^{\dagger}_q]] + \ldots
\end{equation}
We need to evaluate the commutators $\big[\hat\kappa,c_q^\dagger\big]$, $\big[\hat\kappa,\big[\hat\kappa,c_q^\dagger\big]\big]$, $\ldots$, for which we get
\begin{align}
	\big[\hat\kappa,c_q^\dagger\big]&=\sum_{q'q''}\big(\log\hat f\big)_{q'q''}\big[c_{q''}^\dagger c_{q'},c_q^\dagger\big]\nonumber\\
	&=\sum_{q'q''}\big(\log\hat f\big)_{q'q''}\big(c_{q''}^\dagger c_{q'} c_q^\dagger-c_q^\dagger c_{q''}^\dagger c_{q'}\big)\nonumber\\
	&=\sum_{q'q''}\big(\log\hat f\big)_{q'q''}\big(\delta_{q'q}c_{q''}^\dagger-c_{q''}^\dagger c_q^\dagger c_{q'}-c_q^\dagger c_{q''}^\dagger c_{q'}\big)\nonumber\\
	&=\sum_{q'}\big(\log\hat f\big)_{qq'}c_{q'}^\dagger
\end{align}
and
\begin{align}
	\big[\hat\kappa,\big[\hat\kappa,c_q^\dagger\big]\big]&=\sum_{q'}\big[\log\hat f\big]_{qq'}\big[\hat\kappa,c_{q'}^\dagger\big]\nonumber\\
	&=\sum_{q'}\big[\log\hat f\big]_{qq'}\sum_{q''}\big[\log\hat f\big]_{q'q''}c_{q''}^\dagger\nonumber\\
	&=\sum_{q''}\big[\big(\log\hat f\big)^2\big]_{qq''}c_{q''}^\dagger,
\end{align}
for the first two, where $\delta$ is the Kronecker delta.  We see that each commutator simply adds a power to $\log \hat{f}$.  Returning to the Baker--Campbell--Hausdorff formula, this gives 
\begin{align}
	\hat U &c_q^\dagger\hat U^\dagger =e^{\hat\kappa}c_q^\dagger e^{-\hat\kappa}\nonumber\\
	&=c_q^\dagger+\big[\hat\kappa,c_q^\dagger\big]+\frac{1}{2!}\big[\hat\kappa,\big[\hat\kappa,c_q^\dagger\big]\big]+...\nonumber\\
	&=\sum_{q'}\big[\hat I+\log\hat f+\frac{1}{2!}\big(\log\hat f\big)^2+...\big]_{qq'}c_{q'}^\dagger\nonumber\\
	&=\sum_{q'}\big[e^{\log\hat f}\big]_{qq'}c_{q'}^\dagger\nonumber\\
	&=\sum_{q'} \hat{f}_{qq'}c_{q'}^\dagger\nonumber\\
	&=a_q^\dagger,
\end{align}
where $\hat I$ is the identity operator. 

Inserting identities into the string of creation operators we arrive at Eq.~\eqref{Uc}: 
\begin{align}
	\hat U\ket{\Phi_n} &=\hat{U} \prod_{q} (c^{\dagger}_q)^{n_q} \ket{0} \nonumber\\
	&=  \prod_{q}\hat{U} (c^{\dagger}_q)^{n_q} \hat{U}^{\dagger} \ket{0}\nonumber\\
	&=  \prod_{q} (a^{\dagger}_q)^{n_q} \ket{0}\nonumber\\
	&=\ket{\Psi_n},
\end{align} 
where we have used $\hat{U}\ket{0} = \ket{0}$.  

\section{Circuit model for the Thouless operator}
\label{CMFTTO}

In order to implement $\hat{U}(\hat{f})$ on the quantum computer, we use single-particle rotations to diagonalize the one-body change-of-basis-operator $\hat{f}$ and then find the Thouless operator for each of those rotations.  We use two basic single-particle rotations:
\begin{equation}
    \begin{split}
         [\hat r^y_{ij}(\phi_y)]_{kl} &= \delta_{kl} + (\cos \phi_y-1)\delta_{ik}\delta_{il}+(\cos \phi_y-1)\delta_{jk}\delta_{jl}
         \\
         &+ \sin \phi_y \delta_{ik}\delta_{jl} - \sin \phi_y \delta_{jk}\delta_{il}
    \end{split}
\end{equation}
and 
\begin{equation}
    [\hat r^z_i(\phi_z)]_{kl} = \delta_{kl} +  (e^{i\phi_z}-1)\delta_{ik}\delta_{il},
\end{equation}
where the index $j$ defines the row in $\hat f$ that we are targeting and index $i$ defines the row we are rotating into.
In matrix form
\begin{equation}
    r^y_{ij}(\phi_y) = \begin{pmatrix}
     \ddots & \vdots & \vdots & \vdots & \ddots & \vdots & \vdots & \vdots & \ddots \\
     \ldots & 1 & 0 & 0 & \ldots & 0 & 0 & 0 & \ldots \\
     \ldots & 0 & \cos \phi_y & 0 & \ldots & 0 & \sin \phi_y & 0 & \ldots \\
     \ldots & 0 & 0 & 1 & \ldots & 0 & 0 & 0 & \ldots \\
     \ddots & \vdots & \vdots & \vdots & \ddots & \vdots & \vdots & \vdots & \ddots \\
     \ldots & 0 & 0 & 0 & \ldots & 1 & 0 & 0 & \ldots \\
     \ldots & 0 & -\sin \phi_y  & 0 & \ldots & 0 & \cos \phi_y & 0 & \ldots \\
     \ldots & 0 & 0 & 0 & \ldots & 0 & 0 & 1 & \ldots \\
     \ddots & \vdots & \vdots & \vdots & \ddots & \vdots & \vdots & \vdots & \ddots \\
    \end{pmatrix}
\end{equation}
and 
\begin{equation}
    r^z_j(\phi_z) = \begin{pmatrix} \ddots & \vdots & \vdots & \vdots & \ddots  \\
    \ldots & 1 & 0 & 0 & \ldots \\
    \ldots & 0 & e^{i\phi_z} & 0 & \ldots \\
    \ldots & 0 & 0 & 1 & \ldots \\
    \ddots & \vdots & \vdots & \vdots & \ddots  
    \end{pmatrix}.
\end{equation}
We use these rotations to rotate specific off-diagonal indices of $\hat f$ to zero.  
The specific element in row $j$ of $\hat f$ that becomes zero depends on the angles $\phi_y$ and $\phi_z$.  Since $\hat f$ is unitary, we only need to worry about the lower-left elements.  The upper-right elements will be automatically eliminated.  In order to avoid undoing an element that has already been eliminated, we start at the lower left corner and work up in the following order:
\begin{equation}
    \begin{pmatrix}
    * & * & * & * & * & * & \\
    5 & * & * & * & * & * & \\
    4 & 6 & * & * & * & * & \\
    3 & 5 & 7 & * & * & * & \\
    2 & 4 & 6 & 8 & * & * & \\
    1 & 3 & 5 & 7 & 9 & * & \\
    \end{pmatrix},
\end{equation}
where the asterisks indicate indices that do not need to be explicitly targeted and indices with the same value can be eliminated in any order.

To avoid complications with fermion exchange, we restrict $r^y_{ij}(\phi)$ so that $j = i + 1$.
In this case, we can compute the Thouless operators for both $r^y_{ij}(\phi)$ and  $r^z_j(\phi)$ and write them in terms of quantum gates  
\begin{equation}
\label{Urz}
     U[r^z_{j}(\phi_z)] = e^{i\phi_z c^{\dagger}_jc_j} = R^Z_j(\phi_z) 
\end{equation}
up to a global phase and
\begin{equation}
\label{Ury}
\begin{split}
    U[r^y_{ii+1}(\phi_y)] &= e^{\phi_y (c^{\dagger}_ic_{i+1} - c^{\dagger}_{i+1}c_i)} = R^{XY}_{i,i+1}(\phi_y)R^{YX}_{i,i+1}(-\phi_y),
\end{split}
\end{equation}
where 
\begin{equation}
    R^Z_j(\phi) = e^{-i \frac{\phi}{2} Z_{j}},
\end{equation}
is a rotation about the z-axis of the Bloch sphere of qubit $j$, and 
\begin{equation}
\begin{split}
    R^{XY}_{ij}(\phi) = e^{-i \frac{\phi}{2} X_iY_j} ,  \\
    R^{YX}_{ij}(\phi) = e^{-i \frac{\phi}{2} Y_iX_j} ,
\end{split}
\end{equation}
are two qubit rotations.  The two qubit rotations can be generated by single-qubit rotations and C-not gates
\begin{equation}
\begin{split}
    R^{XY}_{ij}(\phi) = R^{Y}_{i}(\pi/2)C^X_{ij}R^Y_j(\phi)C^X_{ij}R^Y_i(-\pi/2),   \\
    R^{YX}_{ij}(\phi) = R^{Y}_{j}(\pi/2)C^X_{ji}R^Y_i(\phi)C^X_{ji}R^Y_j(-\pi/2), 
\end{split}
\end{equation}
where $R^{Y}_{i}(\phi)$ is a rotation about the y-axis of the Bloch sphere of qubit $i$ and $C^X_{ij}$ is a controlled-not gate, targeting qubit $j$ and controlled by qubit $i$.  

Thus, we have a map from the single-particle rotations to quantum gates.  Since we know how to use the single-particle rotations to diagonalize $\hat{f}$, we can reverse the rotations to generate $\hat{f}$ from the identity operator.  Transforming these reverse rotations onto quantum gates allows us to apply $\hat{U}(\hat{f})$ on the quantum computer.  See Appendix~\ref{DS} for a depiction of the circuit model.  

A similar constructions of the quantum gates which generate a Thouless operator can be found in~\cite{Kivlichan2018,Jiang2018} where~\cite{Jiang2018} provides an informative example.  For alternative constructions see~\cite{Wecker2015,Somma2002} .  

\section{Derivation of $\bra{\Psi_0}\hat{G}\hat{H}\hat{G}\ket{\Psi_0}$ }
\label{do}

Due to the idempotency of number operators ($\hat{n}_q^2 = \hat{n}$), we can write the Jastrow-Gutzwiller operator in Eq.~\eqref{J-G} as 
\begin{equation}
    \hat{G}(\theta) = \prod_{q q' }(1+\gamma_{q q'}\hat{n}_q \hat{n}_{q'}),
\end{equation}
where $\gamma_{qq'} = e^{-\theta_{qq'}} - 1$.  To analyze $\hat{G}(\theta)\hat H\hat{G}(\theta)$, we use the identities $\hat{n}_q c^{\dagger}_q = c^{\dagger}_q$, $c_q\hat{n}_q = c_q$, and $c^{\dagger}_q \hat{n}_q = n_q c_q = 0$.  Thus, we can write
\begin{equation}
\begin{split}
    \hat{G}(\theta)\hat H\hat{G}(\theta) =& \sum_{l=0}^{N_H-1} h_l \hat{N}_{\mathcal Q^N_l} C^{\dagger}_{\mathcal Q^+_l} C_{\mathcal Q^-_l}
    \\
    \times& \prod_{qq' \in \bar{\mathcal Q}_l} (1+\gamma_{q q'}\hat{n}_q \hat{n}_{q'} )^2
    \\
    \times& \prod_{q \in \bar{\mathcal Q}_l}\prod_{q' \in \mathcal Q_l  } (1+\gamma_{qq'}\hat{n}_q)
    \\
    \times& \prod_{q \in \mathcal Q_l  }\prod_{q' \in \bar{\mathcal Q}_l} (1+\gamma_{qq'}\hat{n}_{q'} )
    \\
    \times&\prod_{qq'\in \mathcal Q^+_l}(1+\gamma_{qq'})
    \\
    \times&\prod_{qq'\in \mathcal Q^-_l}(1+\gamma_{qq'}).
\end{split}
\label{OHOc2}
\end{equation}
We then convert the number operator factors back into exponentials to arrive at Eq.~\eqref{OHOf} in the main text.

To perform measurements of $\hat{G}(\theta)\hat H\hat{G}(\theta)$ on a quantum computer, we  map the fermion operators onto qubit operators   using the Jordan--Wigner transformation, Eq.~\eqref{J-W}.  After the transformation, we have
\begin{equation}
\begin{split}
    &\hat{G}(\theta)\hat H\hat{G}(\theta) = \sum_{l=0}^{\tilde{N}_H-1} \tilde{h}_l \hat{A}_{\mathcal Q_l}\hat{B}_{\bar{\mathcal Q}_l}
    \\
    &\times \exp\left[-\sum_{q q'}\left( \epsilon_{lq} + \epsilon_{lq'} + \zeta_{lqq'}\right) \theta_{qq'}\left(\frac{1-Z_{q}}{2}\right)^{\epsilon_{lq}}\left(\frac{1-Z_{q'}}{2}\right)^{\epsilon_{lq'}}\right]
\end{split}    
\label{OHOp}
\end{equation}
where $\tilde{N}_H$ is the number of unique Pauli-strings that results from the transformed Hamiltonian, $\hat{A}_{\mathcal Q_l}$ are Pauli-strings containing only Pauli-x and Pauli-y operators acting on orbitals in $\mathcal Q_l$, $\hat{B}_{\bar{\mathcal Q}_l}$ are Pauli-strings containing only Pauli-z  and identity operators acting on orbitals in $\bar{\mathcal Q}_l$, 
and $\epsilon_{lq}$ and $\zeta_{lqq'}$ are defined in Eqs.~\eqref{elq} and \eqref{zlq}.  

All of the Pauli-z operators are diagonal in the $\{\ket{\Phi_n}\}$ basis.    In order to measure term $l$ in Eq.~(\ref{OHOp}), we  rotate the state so that each Pauli-x and Pauli-y operator becomes diagonal.  Let $\ket{\Phi^l_n}$ be the state that has been rotated to measure term $l$ in Eq.~(\ref{OHOp}).  Because none of the Pauli-z operators act on the same index as the Pauli-x or Pauli-y operators, every operator in term $l$ is diagonal in the $\ket{\Phi^l_n}$ basis.  Let $\hat{A}_{\mathcal Q_l} \ket{\Phi^l_m} = \ket{\Phi^l_m}a_{l m}$, $\hat{B}_{\bar{\mathcal Q}_l} \ket{\Phi^l_m} = \ket{\Phi^l_m}b_{l m}$, and $(1-Z_{q\in\bar{\mathcal Q}_l})/2 \ket{\Phi^l_m} = \ket{\Phi^l_m}  n_{q l m} $.  We decompose the final state Eq.~\eqref{Uc} into linear combinations of the measurement basis
\begin{equation}
    \ket{\Psi_n} = \sum_{m=0}^{N_S-1} u^l_{nm} \ket{ \Phi^l_m},
\end{equation}
where $N_S = 2^{Q}$ is the number of many-body basis states.  This yields
\begin{equation}
\begin{split}
    &\bra{\Psi_n}\hat{G}(\theta)\hat H\hat{G}(\theta)\ket{\Psi_n} = \sum_{l=0}^{\tilde{N}_H-1} \tilde{h}_l\sum_{m=0}^{N_S-1}|u^l_{nm}|^2  a_{lm}b_{lm}
    \\
    &\times \exp\left(-\sum_{q q'}\left[ \epsilon_{lq} + \epsilon_{lq'} + \zeta_{lqq'}\right] \theta_{qq'}n_{qlm}^{\epsilon_{lq}}n_{q'lm}^{\epsilon_{lq'}}\right).
\end{split}    
\label{OHOm}
\end{equation}
Each time the quantum computer is measured it collapses into a state $\ket{\Phi^l_m}$ with a random value of $m$ with a probability of $|u^l_{nm}|^2$.  For each state $\ket{\Phi^l_m}$, there are associated values $a_{q l m}$, $b_{q l m}$, and $n_{q l m}$ that are obtained from the measurement result.  We can get an accurate estimate of the statistically significant values of $|u^l_{nm}|^2$ with a subexponential number of shots.  The number of measurements scales only with the number of Pauli strings in the Hamiltonian $\tilde{N}_H$.  As there are at most $Q^2$ non-zero values of $\theta_{qq'}$. the full classical calculation scales as $\tilde{N}_H \times Q^2$.  
Subexponential scaling is achieved by collecting the eigenvalues $n_{qlm}$ into the exponential function Eq.~\eqref{OHOm}.  

\section{Derivation of $\hat{G}(\theta)\hat{K}\hat{G}(\theta)$}
\label{DOGKG}

To derive Eq.\,(\ref{GKG}), we expand $\hat{G}(\theta)$ as
\begin{equation}
\begin{split}
    \hat{G}(\theta) &= e^{-\theta\sum_i n_{i\uparrow}n_{j\downarrow}}
    \\
    &= \prod_ie^{-\theta \hat{n}_{i\uparrow}\hat{n}_{i\downarrow}}
    \\
    &= \prod_i \left( 1 - \theta \hat{n}_{i\uparrow}\hat{n}_{i\downarrow} + \frac{1}{2}\theta^2  \hat{n}_{i\uparrow}\hat{n}_{i\downarrow} - \ldots \right)
    \\
    &= \prod_i \left[1 - \left(\theta - \frac{1}{2}\theta^2 +   \ldots \right)\hat{n}_{i\uparrow}\hat{n}_{i\downarrow} \right]
    \\
    &=  \prod_i\left[1 -  (1-e^{-\theta})\hat{n}_{i\uparrow}\hat{n}_{i\downarrow} \right]   ,
\end{split}
\end{equation}
where in the second line we used $\hat{n}_{i\sigma}^2 = \hat{n}_{i\sigma}$.  From this expression, we can calculate $\hat{G}(\theta)\hat{K}\hat{G}(\theta)$.  First we examine how $\hat{G}(\theta)$ acts on an arbitrary pair of creation and destruction operators:  
\begin{equation}
\begin{split}
    &\hat{G}(\theta)c^{\dagger}_{j\uparrow}c_{k\uparrow}\hat{G}(\theta) 
    \\
    &= \left[\prod_i(1-\lambda \hat{n}_{i\uparrow}\hat{n}_{i\downarrow})\right]c^{\dagger}_{j\uparrow}c_{k\uparrow}\prod_{i'}(1-\lambda \hat{n}_{i'\uparrow}\hat{n}_{i'\downarrow})
    \\
    &=(1-\lambda \hat{n}_{j\uparrow}\hat{n}_{j\downarrow})(1-\lambda \hat{n}_{k\uparrow}\hat{n}_{k\downarrow})c^{\dagger}_{j\uparrow}c_{k\uparrow}
    \\
     &\quad
    \times (1-\lambda \hat{n}_{j\uparrow}\hat{n}_{j\downarrow})(1-\lambda \hat{n}_{k\uparrow}\hat{n}_{k\downarrow})\prod_{i\neq j,k}
     e^{-2\theta\hat{n}_{i\uparrow}\hat{n}_{i\downarrow}}
    \\
    &=  (1-\lambda \hat{n}_{j\downarrow})c^{\dagger}_{j\uparrow}c_{k\uparrow}(1-\lambda \hat{n}_{k\downarrow})\prod_{i\neq j,k} e^{-2\theta\hat{n}_{i\uparrow}\hat{n}_{i\downarrow}}
    \\
    &=  c^{\dagger}_{j\uparrow}c_{k\uparrow}e^{-\theta \hat{n}_{j\downarrow}}e^{-\theta\hat{n}_{k\downarrow}}\prod_{i\neq j,k} e^{-2\theta \hat{n}_{i\uparrow}\hat{n}_{i\downarrow}},    
\end{split}
\label{GccG}
\end{equation}
where $\lambda = e^{-\theta}-1$ and we used $\hat{n}_{i\sigma}c^{\dagger}_{i\sigma} = c^{\dagger}_{i\sigma}$, $c_{i\sigma}\hat{n}_{i\sigma} = c_{i\sigma}$, and $\hat{n}_{i\sigma}c_{i\sigma} = c^{\dagger}_{i\sigma}\hat{n}_{i\sigma} = 0$.  Using Eq.\,(\ref{GccG}) and an analogous one for spin down operators, it is straightforward to calculate $\hat{G}(\theta)\hat{K}\hat{G}(\theta)$:
\begin{equation}
\begin{split}
    &\hat{G}(\theta)\hat{K}\hat{G}(\theta)
    \\
    &= \sum_{\sigma}\sum_{\langle i,j \rangle} \left[ \hat{G}(\theta)c^{\dagger}_{j\sigma}c_{k\sigma}\hat{G}(\theta) + \hat{G}(\theta)c^{\dagger}_{k\sigma}c_{j\sigma}\hat{G}(\theta) \right], 
    \\
    &=\sum_{\sigma }\sum_{\langle i,j \rangle}\left(c^{\dagger}_{j\sigma}c_{k\sigma}+c^{\dagger}_{k\sigma}c_{j\sigma}\right)
    \\
    &\quad\times e^{-\theta \hat{n}_{j\bar\sigma}}e^{-\theta \hat{n}_{k\bar\sigma}}\prod_{i \neq j,k}e^{-2\theta \hat{n}_{i\uparrow}\hat{n}_{i\downarrow}} ,
\end{split}
\end{equation}
where $\bar\sigma$ is the complement of $\sigma$.  With the Jordan--Wigner transformation we arrive at the expression in Eq.~\eqref{GKG}.

\section{Device specifications}
\label{DS}

\begin{figure*}[t]
\vspace{2mm}
\begin{center}
\includegraphics[width=2\columnwidth]{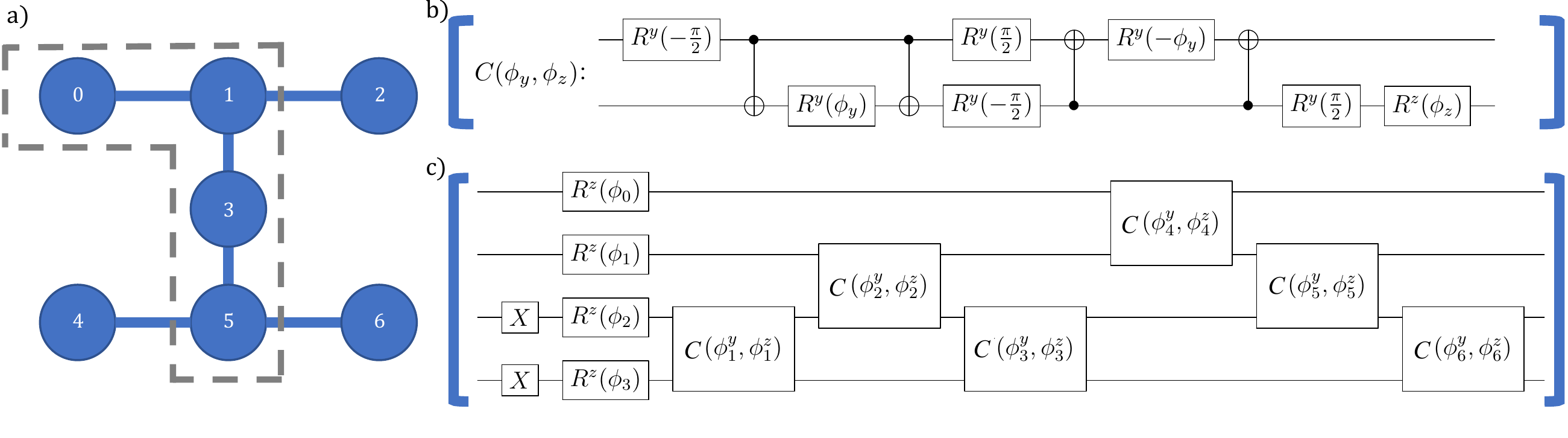}
\end{center}
\vspace{-2mm}
\caption{Circuit model specifications.  a) layout of the IBM Lagos device.  Qubits are depicted as circles and the connectivity is shown by the lines.  The dashed boarder shows the qubits that were used in our demonstration.  b) Circuit model for the operator $C(\phi_y,\phi_z) = U[r_{i,j}^y(\phi_y)]U[r_j^z(\phi_z)]$ as defined in Eq.~\eqref{Urz} and \eqref{Ury}.  The two qubit gates are $C^X$ gates, the $R^y$ gates represent rotations around the $y$-axis of the block sphere, and $R^z$ gates represent rotations around the $z$-axis.  c) circuit model for the full Thouless operator.  From top to bottom the lines are applied to qubit $q_0$, $q_1$, $q_3$, $q_5$ as shown if (a).  The state is set to half filling with the initial Pauli-X gates.  The $R^z$ gates, in the next step, set up the diagonalized $\hat{f}$ matrix.  The $C$ gates are defined in (b).  All angles $\phi_i$, $\phi^y_i$, and $\phi^z_i$ are found during the diagonalization process of $\hat{f}$ as described in Appendix~\ref{CMFTTO}.   
}
\label{circuit}
\vspace{-3mm}
\end{figure*}

We demonstrate our method on IBM Q Lagos which is a seven-qubit quantum computer.  Lagos has the following qubit pairs connected
\begin{equation}
    \begin{matrix}
           & q_0 & q_1 & q_2 & q_3 & q_4 & q_5 & q_6 \\
     q_0   & 0   & 1   & 0   & 0   & 0   & 0   & 0   \\
     q_1   & 1   & 0   & 1   & 1   & 0   & 0   & 0   \\
     q_2   & 0   & 1   & 0   & 0   & 0   & 0   & 0   \\
     q_3   & 0   & 1   & 0   & 0   & 0   & 1   & 0   \\
     q_4   & 0   & 0   & 0   & 0   & 0   & 1   & 0   \\
     q_5   & 0   & 0   & 0   & 1   & 1   & 0   & 1   \\
     q_6   & 0   & 0   & 0   & 0   & 0   & 1   & 0 & ,   \\
    \end{matrix}
\end{equation}
where $q_i$ refers to qubit $i$ and $1$/$0$ specifies pairs which are connected/disconnected. The connectivity map of Lagos is shown if Fig~\ref{circuit}a.  We used only qubits $q_0$, $q_1$, $q_3$, and $q_5$.    The pre-transpilation circuit is shown in Fig~\ref{circuit}(b,c).  The circuit in Fig~\ref{circuit}b represents a single step during the diagonalization process of $\hat{f}$ as discussed in Appendix~\ref{CMFTTO}.  This step is repeated several times with different angles and applied to different qubits during the full implementation of the Thouless operator in Fig.~\ref{circuit}c.    There are 24 CX gates in the circuit which is the largest contributing factor to the error.  After transpilation, all of the gates were converted to z-rotations $R^z$, root-x gates $\sqrt{X}$, and controlled-not gates CX using the qiskit transpile function.  The specifications of the device including the error rates of each gate can be found in Table~\ref{T1}.  These specifications are based on a calibration performed on May 5, 2022.    The $R^z$ gates have no error as they are implemented by changing the measurement angles.  

\begin{table}[h]
\caption{\label{tab:example} Specifications of IBM Q Lagos during the execution of our algorithm.}
\begin{ruledtabular}
\begin{tabular}{lccccccc}
 & $q_0$ & $q_1$ & $q_2$ & $q_3$ & $q_4$ & $q_5$ & $q_6$ \\
Frequency (GHz)  & 5.24 & 5.1 & 5.19 & 4.99 & 5.29 & 5.18 & 5.06 \\ T1 ($\mu$s)   & 137.3 & 136.2 & 179.4 & 152.2 & 102.7 & 134.4 & 142.6 \\ T2 ($\mu$s)   & 49.8 & 114.6 & 130.1 & 76.4 & 44.9 & 98.2 & 194.0 \\ Readout error (\%)  & 0.97 & 0.71 & 1.42 & 1.78 & 1.41 & 2.78 & 0.72 \\ $\sqrt{X}$ error (\%)   & 0.018 & 0.023 & 0.022 & 0.018 & 0.036 & 0.063 & 0.028 \\ $\text{CX}_{q_0,q_i}$ error (\%)    &  & 0.707  &   &   &   &   &  \\$\text{CX}_{q_1,q_i}$ error (\%)  & 0.707  &  & 0.661 & 0.675  &   &   &  \\$\text{CX}_{q_2,q_i}$ error (\%)   &  & 0.661  &   &   &   &   &  \\$\text{CX}_{q_3,q_i}$ error (\%)    &  & 0.675  &   &   &  & 1.163  &  \\$\text{CX}_{q_4,q_i}$ error (\%)    &   &   &   &   &  & 3.406  &  \\$\text{CX}_{q_5,q_i}$ error (\%)   &   &   &  & 1.163 & 3.406  &  & 0.908 \\$\text{CX}_{q_6,q_i}$ error (\%)   &   &   &   &   &  & 0.908  &  \\
\end{tabular}
\end{ruledtabular}
\label{T1}
\end{table}

\clearpage

\bibliographystyle{apsrev4-2}
\bibliography{REF}


\end{document}